# A large dataset of human EEG responses to short naturalistic videos for studying dynamic visual event processing

Alessandro T. Gifford[1,2,3,*], Pablo Oyarzo[1],
Anne W. Zonneveld[4], Christina Sartzetaki[4],
Iris I.A. Groen[4], Radoslaw M. Cichy[1,2,3,5]

[1] Department of Education and Psychology, Freie Universität Berlin, Berlin, Germany
[2] Einstein Center for Neurosciences Berlin, Charité – Universitätsmedizin Berlin, Berlin, Germany
[3] Bernstein Center for Computational Neuroscience, Humboldt-Universität zu Berlin, Berlin, Germany
[4] Informatics Institute, University of Amsterdam, Amsterdam, The Netherlands
[5] Berlin School of Mind and Brain, Humboldt-Universität zu Berlin, Berlin, Germany

* Correspondence: alessandro.gifford@gmail.com

# Abstract

Vision neuroscience has experienced a surge in the collection and use of large-scale datasets of brain responses to naturalistic images. However, static images lack the temporal dimension essential for understanding how vision is solved in the brain during dynamic real life settings. To facilitate the study of the neural correlates of dynamic visual event perception, we introduce the EEG Moments Dataset (EMD). EMD consists of 128-channel EEG responses and eye-tracking recordings of 6 human participants viewing 1,102 short naturalistic videos (3-second long; with audio track) while maintaining central fixation. We show that EMD's EEG responses well encode stimulus-related information, exhibit a temporal correspondence with the video stimuli, and have a rich representational content revealed by brain encoding models based on different feature spaces. Furthermore, complemented by the BOLD Moments Dataset (BMD) – an existing large-scale dataset of human functional magnetic resonance imaging (fMRI) responses for the same videos – EMD enables spatio-temporally resolved investigations of brain responses to dynamic visual events. We release EMD's EEG and eye-tracking data in both raw and preprocessed format, along with the 1,102 video stimuli, and rich stimulus metadata. Finally, we provide an interactive code tutorial to familiarize with EMD's preprocessed data, stimuli, and stimulus metadata.

# Background & Summary

We introduce a large-scale dataset of electroencephalography (EEG) responses called EEG Moments Dataset (EMD). EMD consists of dense 128-channel EEG recordings of 6 human participants viewing 1,102 short naturalistic videos (3-second long; with audio track) while maintaining central fixation, and of eye-tracking data (gaze position and pupil size) recorded throughout video presentation. This extensive sampling of single participants aims at facilitating computational visual neuroscience research, which hinges on large amounts of data for model development and detailed investigation of the brain[1,2]. Moreover, the high temporal resolution of EMD's EEG responses complements the high spatial resolution of the BOLD Moments Dataset (BMD)[3] – an existing large-scale dataset of functional magnetic resonance imaging (fMRI) responses of 10 participants for the same 1,102 video stimuli – thus enabling spatio-temporally resolved analyses[4–6] of dynamic visual event processing in the human brain. We provide EEG and eye-tracking data in both raw and preprocessed format, the 1,102 video stimuli, and rich stimulus metadata including object labels, scene labels, action labels, text descriptions, spoken transcriptions, memorability scores, and motion energy features. We release all code used to collect, preprocess, and validate EMD, and we additionally provide an interactive code tutorial in Python to familiarize with EMD's stimuli and stimulus metadata, as well as with the preprocessed EEG and eye-tracking data.

The focus on short (3 second) naturalistic videos is motivated by the higher ecological validity compared with naturalistic static images (which have been a popular choice for recent large-scale visual neural datasets[7,8,5,9,10]), while at the same time retaining experimental control through temporally bounded events that isolate the phenomena of interest from prolonged contextual effects[11]. Compared to images, videos have a temporal dimension which provides unique contextual information about how spatial components in our environment move, change, and spatially relate to each other over time (e.g., the train door is opening, and not closing), or about how events unfold over time (e.g., a person getting emotional only after, and not before, reading a note). Thus, the increase of stimulus ecological validity from static images to videos supports understanding how vision is solved in the human brain in dynamic real life settings.

We provide foundational analyses that thoroughly validate EMD for computational visual neuroscience research. These analyses show that EMD is well suited for studying neural processing of dynamic visual events in the human brain at both high temporal and, in combination with BMD, also high spatial resolution. First, through event-related potential (ERP), signal-to-noise, and decoding analyses, we show that EMD's EEG responses encode stimulus-related information. These results revealed dynamics of representational content beyond those previously revealed for static images[4,8,12,13]. Second, through computational modeling analyses we show a temporal correspondence between the dynamics of EMD's video stimuli and the time courses of the corresponding EEG responses, indicating that EMD enables fine-grained analyses of how dynamic visual events unfold across the EEG time course. Third, through deep-neural-network-based encoding models[14] we show that EMD's EEG responses have a rich representational content related to a diverse set of stimulus feature spaces (i.e., vision- or language-based stimulus features), indicating that EMD is well poised for computational modeling through deep-learning-based approaches. (This is further demonstrated by a recent study that leveraged EMD's rich representational content for a large-scale modeling effort[15].) Fourth, we show that combining EMD's EEG data with BMD's fMRI data enables a spatio-temporally resolved investigation of brain responses to dynamic visual events.

# Methods

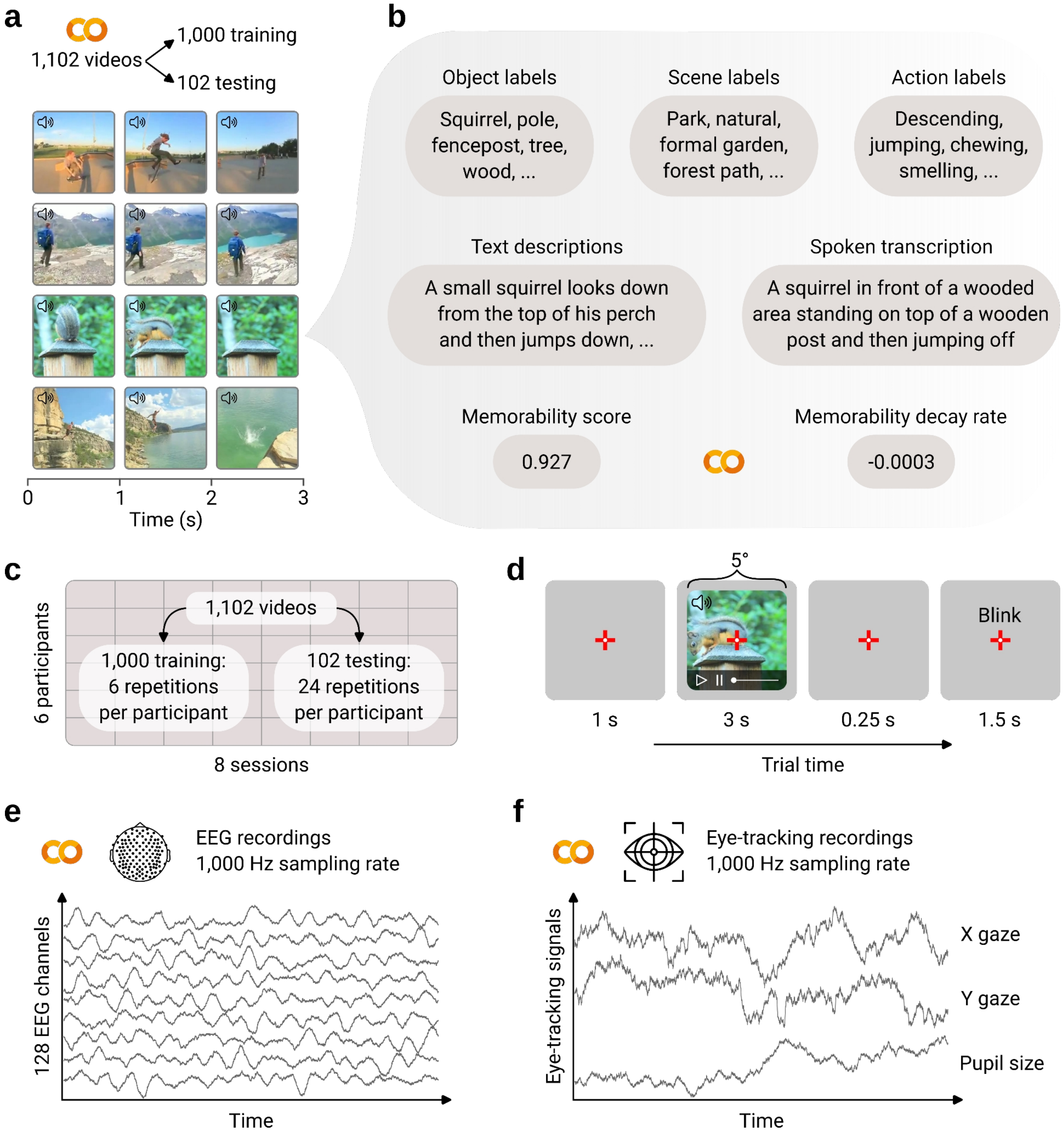

**Fig. 1** EMD's stimuli, stimulus metadata, experimental design, and data acquisition. (**a**) EMD's stimuli consist of 1,102 short naturalistic videos (3-second long), of which 1,000 belong to the dataset training set, and 102 to the dataset testing set. These are the same videos used in the BOLD Moments Dataset[3] (BMD), with the difference that we also presented the video audio track during the EMD experiment. (**b**) Each video stimulus has rich metadata including object labels, scene labels, action labels, text descriptions, a spoken transcription, a memorability score and decay rate, and motion energy features (not shown here). All metadata come from BMD[3]. (**c**) EMD consists of 6 participants with 8 data collection sessions each. During these 8 sessions, each of the 1,000 training videos were repeated 6 times, and each of the 102 testing videos were repeated 24 times. (**d**) Each experimental trial started with 1 s of blank screen, followed by 3 s of video presentation, 0.25 s of blank screen, and ended with 1.5 s where participants were instructed to blink. To ensure that participants attended to the videos, during ~9% of randomly chosen trials per run the videos were followed by one object, scene, or action label, and participants were asked to report whether this label matched the video content. Videos were presented on a gray background at the center of the screen subtending 5° of visual angle and overlaid with a central red fixation cross (0.52° of visual angle). Participants were instructed to focus on the fixation cross for the duration of the experiment. (**e**) We recorded neural responses throughout the video viewing experiment using 128-channel EEG at a sampling rate of 1,000 Hz. (**f**) We recorded monocular eye-tracking data throughout the video viewing experiment, including x gaze, y gaze, and pupil size, at a sampling rate of 1,000 Hz. (**a-b, e-f**) We provide an interactive Google Colab code tutorial to familiarize with EMD's stimuli and stimulus metadata, as well as with the preprocessed EEG and eye-tracking data.

## Participants

Six healthy adults (three males aged 21, 28, 32; three females aged 21, 23, 24) participated, all having normal or corrected-to-normal vision. They all provided informed written consent and received monetary reimbursement. Procedures were approved by the ethical committee of the Department of Education and Psychology at Freie Universität Berlin and were in accordance with the Declaration of Helsinki.

## Stimuli

The stimulus set consisted of the same 1,102 videos used in the BOLD Moments Dataset[3] (BMD; parts of this section, and of the stimulus metadata section below, are reproduced from the BMD publication, where the stimuli and stimulus metadata were originally described). However, differently from the BMD experiment, we also used the video's corresponding audio track, so as to enable analyses of visual and/or auditory processing of naturalistic dynamic events (**Fig. 1a**; 34 videos did not contain an audio track and were therefore presented silently; users are invited to familiarize themselves with EMD's video stimuli through the code tutorial accompanying the dataset). The videos were sampled from the Memento10k dataset[21], which is a subset of the Moments in Time dataset[22] and Multi-Moments in Time dataset[23]. Each video was square-cropped and resized to 268×268 pixels. Videos had a duration of 3 seconds and frame rates ranging from 15 to 30 frames per second (mean = 28.3). The 1,102 videos were manually selected from the Memento10k dataset by two human observers to encompass videos that contained movement (i.e., no static content), were filmed in a natural context, and represented a wide selection of possible events a human might witness. Additional criteria were to be free of post-processing effects, textual overlays, excessive camera movements, blur, and objectionable or inappropriate content.

The 1,102 videos selected for the main experiment were split into a training and a testing set; 102 videos were chosen for the testing set, and the remaining 1,000 videos formed the training set. Specifically, the testing set videos were chosen randomly from the 1,102 videos, and then checked manually to ensure no semantic overlap occurred between any pair of testing set videos, in terms of objects plus actions. If semantic overlap was found between a pair of videos as determined by an author, one of these videos was swapped with a video randomly selected from the pool of remaining videos and incorporated into the testing set. This was repeated until a semantically-diverse testing set was formed. The training and testing sets are intended to mirror training and testing sets common in machine learning applications, reflecting potential use for model building and evaluation.

## Stimulus metadata

Visual events consist of complex combinations of objects, locations, actions, and other kinds of content. To capture key dimensions of visual events, each video has the same set of seven metadata categories created for the BOLD Moments Dataset[3] release: object labels, scene labels, action labels, text descriptions, a spoken transcription, a memorability score and an index of memorability decay rate, and motion energy features (**Fig. 1b**; users are invited to familiarize themselves with EMD's stimulus metadata through the code tutorial accompanying the dataset). Five object, scene, action, and text description labels were collected for each stimulus to ensure comprehensive coverage (the five annotators' labels reflect their unique interpretations) and form a group consensus (the five annotators' labels can be used to converge on a single label). Labels for each metadata category were collected in different human crowd-sourced experiments. An annotator was allowed to label up to 20 different videos but no more than one label per video to encourage a diverse sampling of human annotators within videos and throughout the stimulus set. While crowd-source workers were not restricted from participating in multiple metadata experiments, this was unlikely due to the experiments being collected at different times and the large population pool from which the crowd-source workers were drawn.

*Object labels.* For each video, at least 5 sets of up to three different object labels were obtained in a human crowd-sourced experiment on Prolific. Each annotator was instructed to select up to three different object labels visible in the video. They selected at least one object label, and if they believed no more objects were present in the video, they were allowed to select, up to two times, an option labeled "No more objects in the video". This option thus encouraged accurate labels and carried information on the density of objects in the video. Each object label was one of 1,854 possible labels from the THINGS dataset[24] to encourage overlap with computational neuroscience work and leverage the additional THINGS metadata on each label (e.g., animacy, size, indoor/outdoor). The object label selections can be different or the same across annotators. One author manually reviewed the labels to ensure the labels assigned to the video were sensical (i.e., participants were not choosing labels at random).

*Scene labels.* For each video, at least 5 scene labels were obtained using a human crowd-sourced experiment on Prolific. Each of the five different annotators were instructed to select a scene label that best describes the scene of the video. All scene labels came from the Places365 dataset[25] for its broad scene coverage and overlap with computer vision resources. The scene label selections can be different or the same across videos. One author manually reviewed the labels to ensure the labels assigned to the video were sensical (i.e., participants were not choosing labels at random).

*Action labels.* The 5 action labels were selected by workers on the crowd-sourcing platform Prolific. The possible action labels were restricted to be from one of 292 possible action labels that broadly encompass meaningful human actions[23]. The participant viewed these 292 possible action labels, watched the video, and selected one action label that best described the video. Each of the 5 action labels per video were produced by different participants. Given that an action, by definition, unfolds across time and the EMD stimuli have a short 3 second duration, the majority of the videos contain one primary action. In the edge case that a complex video captures multiple salient actions, the five annotations can capture these multiple actions. Note that annotators labeled up to 3 objects in a video (described above) because objects, unlike actions, have clear spatial boundaries, often occur in multiple instances in a video, and have no temporal dimension. Two authors manually reviewed the labels to ensure the labels assigned to the video were sensical (i.e., participants were not choosing labels at random).

*Text descriptions.* Five high-level descriptions of the content of each video were obtained in the form of written descriptions. The five text descriptions were human-generated from participants on the crowd-sourcing platform Amazon Mechanical Turk (AMT). Their task was to watch the video and type a 10–15 word caption in complete English sentences. Each of the 5 text descriptions were produced by a different human participant, and each participant was allowed to annotate multiple videos. The authors manually checked the text descriptions to ensure the text descriptions pertained to the video (i.e., participants followed instructions) and to correct obvious typos.

*Spoken transcriptions.* One spoken description per video was obtained to capture emotional and descriptive nuances typically conveyed in speech but not present in typing. The spoken description was collected via human participants on AMT, as in Monfort and colleagues[26]. Participants were instructed to watch the video and verbally describe it. They were given no instructions pertaining to the length of the description. The authors recorded the audio file and used Google's speech-to-text transcription to generate a text transcription. The transcription was manually checked to ensure it pertained to the video (i.e., participants followed instructions) and to correct obvious typos. For privacy purposes, only the text transcriptions, but not the original audio files, are released.

*Memorability score and decay rate.* The memorability score and decay rate were measured by Newman and colleagues[21], where AMT human participants played a video memory game. The game consisted of a continuous video stream where the participant pressed the spacebar upon seeing a repeated video. Repeated videos were presented at various delays, from 30 seconds to ten minutes. The participant's responses were then used to calculate a video's memorability score from 0 (no recall) to 1 (perfect recall) and memorability decay rate from 0 (no decay) to -inf (instantaneous decay). Specifically, a video's memorability score is the fraction of correct identifications in the memory game normalized to a lag of 80 videos in between consecutive presentations. A video's memorability decay rate describes how a video's memorability score changes over different lags. The memorability decay rate was regressed from the output of SemanticMemNet[21], an Inflated 3D (I3D) Convolutional Neural Network trained to predict a video's memorability score and text caption from the video frames and optical flow. A video's memorability score ($m$) and decay rate ($\alpha$) can be used to predict its memorability at any time lag $t$ (the number of videos in between the first and second presentation) according to the following equation[21]:

$$m_t = m_T + \alpha(t - T)$$

where $T$ is a set lag of 80. The memorability scores given in the stimuli metadata were computed at a lag of $t = 80$.

Note that Newman and colleagues[21] experimentally observed memorability decay rates above 0. Since a positive memorability decay rate may reflect a currently unknown stimulus or cognitive feature, they are preserved in the stimulus metadata. One may wish to clip them to a maximum value of 0 depending on the analysis.

*Motion energy features.* Video-computable motion features of each video were created using a motion energy model[27–29]. The motion energy model uses a set of spatial and temporal Gabor filters to extract a video's motion and direction.

## Experimental design

Each of the 6 participants underwent 8 data collection sessions. During each session, participants were presented with one quartile of the training videos (i.e., 250 out of 1,000 videos), and with all 102 testing videos, each repeated for 3 times. Thus, across all 8 data collection sessions, the 1,000 training videos were repeated 6 times, and the 102 testing videos were repeated 24 times (**Fig. 1c**). The videos were presented in random order with the constraint of no consecutive repetition.

One session comprised 16 runs, each lasting around 7 minutes and consisting of 66 video presentation trials. Each trial started with 1 s of blank screen, followed by 3 s of video presentation, 0.25 s of blank screen, and ended with 1.5 s where participants were instructed to blink or to make any other movement (this reduced the chances of eye blinks and other artifacts during the video presentations). Videos were presented on a gray background at the center of the screen subtending 5° of visual angle and overlaid with a central red fixation cross (0.52° of visual angle). Participants were instructed to focus on the fixation cross for the duration of the experiment (**Fig. 1d**).

To ensure that participants attended to the videos, during 6 random trials per run participants were presented with one object, scene, or action label sampled from the stimulus annotations (see "Stimulus metadata" section above), instead of the 1.5 s blink period at the end of the run. Participants had up to 5 s to report, with a keypress, whether this label matched the video content (the labels matched the video content half of the times). During these up to 5 seconds of task participants were also asked to blink.

Videos were presented using a monitor with a resolution of 1,680 pixels × 1,050 pixels and a refresh rate of 60 Hz. The size of the full monitor image was 47.5 cm (width) × 29.7 cm (height). The audio track of each stimulus was played along the video presentation. We controlled stimulus presentation using the Psychtoolbox[30], and recorded EEG and eye-tracking data during the experimental sessions.

## EEG data acquisition and preprocessing

We recorded the EEG data using a 128-channel Easycap actiCAP snap cap with Brain Products actiCAP slim active electrodes arranged in accordance with the extended 5-10 electrode system[31], and a BrainVision actiCHamp Plus amplifier. We recorded the data at a sampling rate of 1000 Hz, while applying an online low-pass filter with a cutoff frequency of 280 Hz, and referencing to the Fz electrode (**Fig. 1e**). Stimulus-onset triggers sent via a parallel port from the stimulation computer were recorded by the EEG recording system. For some participants, a small number of EEG trials were not recorded due to data collection issues (see "Data Records" section below).

We performed offline preprocessing in Python, using the MNE package[32]. We filtered the data between 0.1 Hz and 40 Hz, ran an automatic identification and removal of blink components based on independent component analysis, and epoched the continuous EEG data into trials ranging from 0.1 s before stimulus onset to 3.5 s after stimulus onset. We then downsampled the data to 500 Hz, ran an automatic rejection of bad trials and interpolation of bad channels using Autoreject[33] with default parameters, and applied baseline correction by subtracting the mean of the pre-stimulus interval for each trial and channel separately. The percentage of retained EEG data after trial rejection was on average 93,91 % across the six participants (**Table 1**). Users are invited to familiarize themselves with EMD's preprocessed EEG data through the code tutorial accompanying the dataset.

## Eye-tracking data acquisition and preprocessing

We recorded monocular eye-tracking data using an EyeLink 1000 system (SR Research). We recorded eye-tracking data (x gaze, y gaze, pupil size) for the right eye, at a sampling rate of 1,000 Hz, and using the Pupil-CR centroid mode (**Fig. 1f**). Before the beginning of each experimental run, we performed a 9-point calibration procedure, followed by a validation stage (with acceptance criteria of worst point error < 1°, and an average error < 0.5°). Stimulus-onset triggers sent via a parallel port from the stimulation computer were recorded by the eye tracker (these triggers were sent simultaneously to the ones sent to the EEG recording system described above, so as to make it possible to synchronize stimulus presentation, EEG data, and eye-tracking data). For some participants, a small number of eye-tracking trials were not recorded due to data collection issues (see "Data Records" section below).

We implemented the following preprocessing steps. First, we removed blinks by setting to NaN samples 150 ms before and 150 ms after each blink occurrence. We then epoched the continuous eye-tracking data into trials ranging from 0.1 s before stimulus onset to 3.5 s after stimulus onset (analogously to the EEG). Next, we converted the gaze data from distance from the left (x gaze) or top (y gaze) screen edges in pixels, to deviation from the center of the screen in degrees of visual angle, and baseline corrected the pupil size data by subtracting the mean of the pre-stimulus interval for each trial separately. We then set to NaN data with a deviation from central fixation of more than 4° of visual angle (i.e., eye movements beyond the stimulus boundaries), excising samples 20 ms before and 20 ms after each occurrence. Finally, we downsampled the data to 500 Hz, and only retained trials with at least 75% valid samples (i.e., non-NaN samples). The percentage of retained eye-tracking data after trial rejection was on average 94,76 % across the six participants (**Table 1**). Users are invited to familiarize themselves with EMD's preprocessed eye-tracking data through the code tutorial accompanying the dataset.

| Participant | EEG | | Eye tracking | |
|---|---|---|---|---|
| | **Total trials** | **Retained trials** | **Total trials** | **Retained trials** |
| 1 | 8,447 | 8,065 (95.48 %) | 8,448 | 8,448 (100 %) |
| 2 | 8,448 | 8,056 (95.36 %) | 8,448 | 8,239 (97.53 %) |
| 3 | 8,448 | 8,067 (95.49 %) | 8,448 | 7,366 (87.19 %) |
| 4 | 8,448 | 7,900 (93.51 %) | 8,345 | 7,824 (93.76 %) |
| 5 | 8,446 | 7,986 (94.55 %) | 8,432 | 8,327 (98.75 %) |
| 6 | 8,448 | 7,525 (89.07 %) | 8,448 | 7,713 (91.3 %) |
| All | 50,685 | 47,599 (93,91 %) | 50,569 | 47,917 (94,76 %) |

**Table 1** Total EEG and eye-tracking trials, and amount of retained trials after preprocessing. For some participants, a small number of EEG and/or eye-tracking trials are not available due to data collection issues, thus resulting in a different number of total trials between participants and/or data modalities.

# Data Records

## Raw data

The EEG Moments Dataset is freely available on OpenNeuro: https://openneuro.org/datasets/ds008257. We followed the BIDS-EEG[34,35] standard for file structures and naming conventions. The main directory contains the raw EEG and eye-tracking data for each subject and session under the “eeg” datatype directory. The eye-tracking data is denoted as “physio” in the corresponding file names. For some participants, a small number of EEG and/or eye-tracking trials are not available due to data collection issues; we list these missing trials in the main README file of the OpenNeuro dataset repository.

## Derivatives

The “derivatives” directory contains the preprocessed EEG and eye-tracking data within the “eeg” and “eyetracking” subdirectories, respectively. These are the data used in the technical validation analyses of this paper. The preprocessed EEG data is stored as separate “./eeg/sub-0X/sub-0X_ses-0Y_preprocessed_eeg.h5” files (*X* indicating the subject number ranging from 1 to 6, and *Y* indicating the session number ranging from 1 to 8), where each file consists of an array of shape (Trials, Channels, Times). In addition, each subject has a “./eeg/sub-0X/sub-0X_eeg_metadata.npy” file consisting of EEG metadata: the stimulus ID, run number, and trial number associated with the EEG trials of each session; the EEG channel names; the time points (in seconds); the noise ceiling signal-to-noise-ratio (NCSNR) and noise ceiling scores[7] estimated on each EEG channel and time point using the preprocessed EEG responses for the 102 testing videos (where the noise variance is the variance across repeated trials, the signal variance is the difference between the total variance and the noise variance, the NCSNR is the ratio of signal standard deviation over noise standard deviation, and the noise ceiling is the ratio of signal variance over total variance indicating the percentage of explainable variance given noise in the data). The preprocessed eye-tracking data is stored as separate “./eyetracking/sub-0X/sub-0X_ses-0Y_preprocessed_eyetracking.h5” files for every subject and recording session, where each file consists of an array of shape (Trials, Signals, Times), with “Signals” including x gaze coordinates, y gaze coordinates, and pupil size. In addition, each subject has a “./eyetracking/sub-0X/sub-0Y_eyetracking_metadata.npy” file consisting of eye-tracking metadata: the stimulus ID, run number, and trial number associated with the eye-tracking trials of each session; the time points (in seconds). Note that the preprocessed EEG and eye-tracking data are not equivalent in number of trials, since during preprocessing EEG and eye-tracking trials were discarded based on independent criteria (see Methods section). However, it is possible to match trials that were kept for both data modalities using the trials’ run and trial number present in the metadata. The “derivatives” directory additionally contains stimulus metadata within the “stimuli_metadata” subdirectory, consisting of annotations, motion energy features, and a list of file names for the 34 videos that did not have an audio track.

## Stimuli

The 1,102 video stimuli are in “.mp4” format. The first 1,000 videos belong to the dataset train split, whereas the remaining 102 videos belong to the dataset testing split. Due to copyright permission considerations, the video stimuli (including audiotrack) must be downloaded separately: the “stimuli” directory contains a “stimuli_download.txt” file with download instructions.

# Technical Validation

All technical validation analyses are based on the preprocessed EEG and eye-tracking data versions prepared following the steps described in the Methods section. Since EMD's EEG data span several recording sessions, they are affected by session-specific effects, that is, differences in EEG responses across sessions due to incidental session-specific factors (e.g., time of day, arousal, cognitive state, hardware state). Thus, to reduce spurious distributional differences due to session-specific effects, we $z$-scored the EEG responses of each channel and time point, across the trials of each recording session independently. We conducted this step prior to all analyses with the exception of the event-related potential (ERP) analysis, since independent $z$-scoring of each time point would destroy meaningful changes in voltage across time points.

## EMD's participants maintained central fixation and their pupil sizes varied based on stimulus luminance

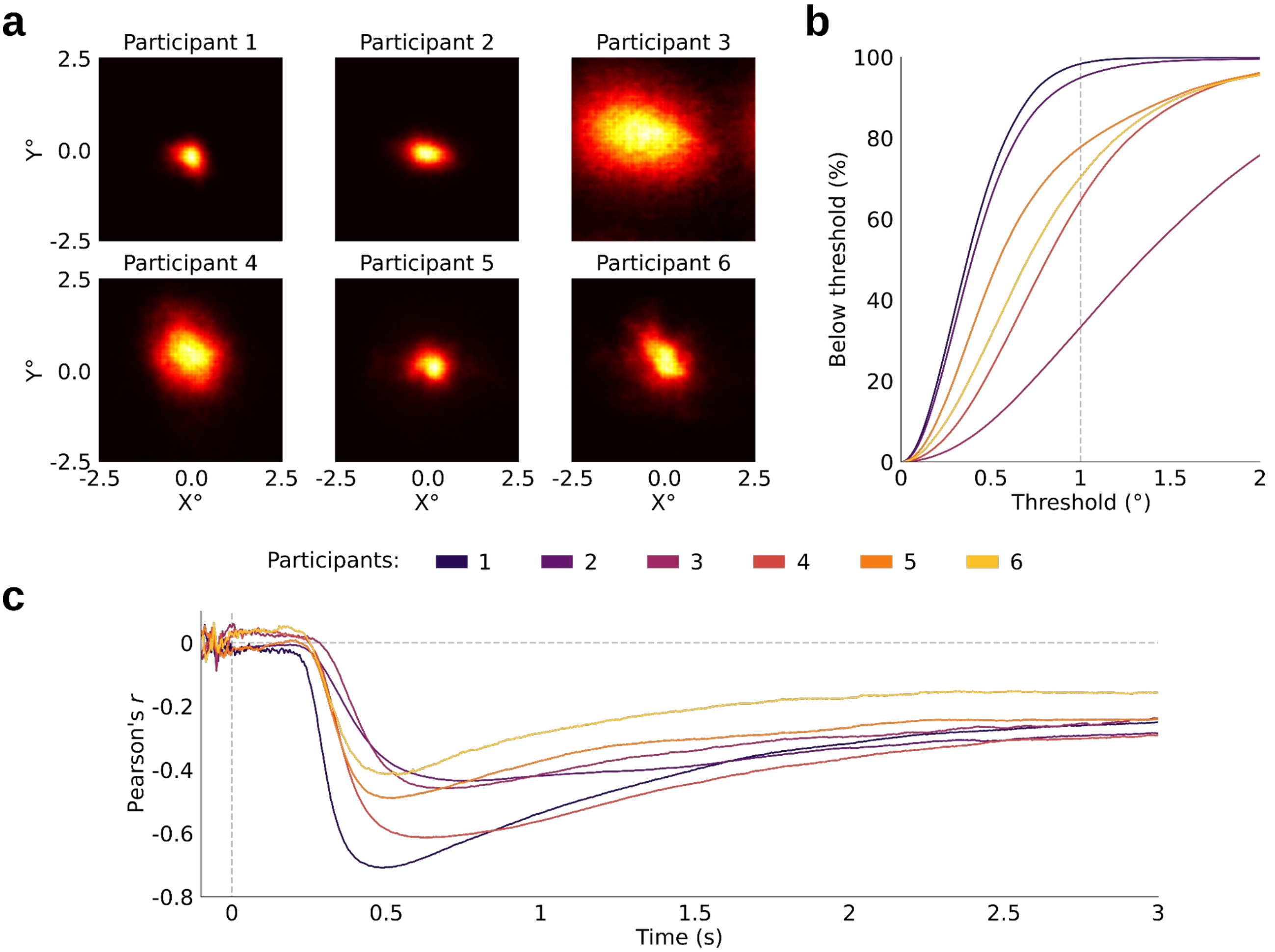

**Fig. 2** EMD's participants maintained central fixation and their pupil sizes varied based on stimulus luminance (**a**) Participant-wise 2D histograms of gaze positions, in visual degrees. (**b**) Participant-wise percentage of time during which deviation from central fixation (*y* axis) was less than a specific visual degree threshold (*x* axis). Results are shown for a range of thresholds. (**c**) Participant-wise correlation (Pearson's *r*) between pupil size and video stimulus luminance.

We first analyzed EMD's eyetracking data by asking whether the participants maintained central fixation during video viewing, and whether their pupil size varied as a function of stimulus luminosity[36]. To assess central fixation, we computed the 2D histograms of gaze position across the 3 s presentation intervals of all videos (**Fig. 2a**), and further computed the cumulative distribution function (CDF) of visual degree deviation from central fixation (**Fig. 2b**). With the exception of participant 3, gaze position was within 1° from central fixation 65-98% of the time, indicating that participants were overall able to maintain fixation. Participant 3 exhibited more substantial eye movements, which are likely the result of calibration difficulties associated with wearing glasses, rather than actual eye movements per se. To assess the relationship between pupil size and stimulus luminance, we correlated the pixel- and frame-averaged luminance of all video stimuli, with the repetition-averaged pupil sizes for the same stimuli, independently for each eye-tracking time point. This resulted in negative correlations starting from 0.3 s after stimulus onset (**Fig. 2c**), in line with previous work demonstrating that pupil diameter increases (or decreases) with decreases (or increases) in stimulus luminance[36]. The fact that pupil size was not entirely explained by stimulus luminance – the strongest correlation scores ranged between -0.4 and -0.7 across participants – invites the investigation of other stimulus attributes determining pupil size such as color, spatial structure, motion, saliency, or emotional arousal[37–42].

EMD’s EEG responses well encode stimulus-related information

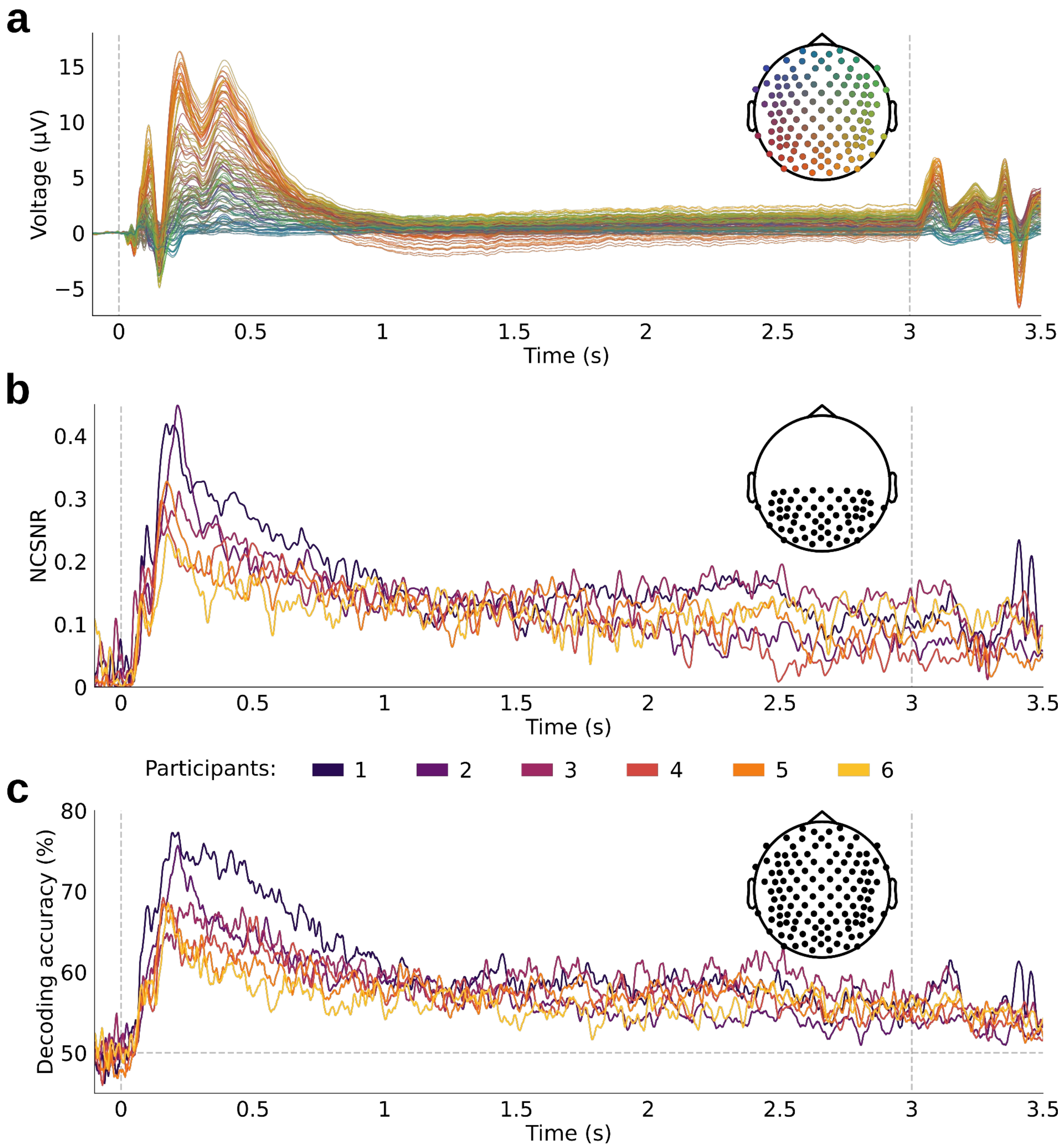

**Fig. 3** EMD’s EEG responses well encode stimulus-related information. (**a**) EEG event-related potentials (ERPs) of participant 1 (the other participants resulted in qualitatively similar ERPs). Each curve corresponds to the ERP of a different channel. (**b**) Participant-wise EEG noise ceiling signal-to-noise ratio (NCSNR) scores, averaged across occipital and parietal EEG channels. (**c**) Participant-wise EEG pairwise decoding scores, obtained from support vector machines (SVMs) trained and tested on EEG responses from all channels.

We next moved to EMD’s EEG responses, and tested whether they contain stimulus-related information through three incremental analyses that revealed common patterns across their results. First, we computed the event-related potentials (ERPs) by averaging the channel-wise preprocessed EEG responses across all trials from all video conditions. The resulting ERPs showed a change in EEG voltage compared to the pre-stimulus baseline, indicating the encoding of stimulus presentation in EMD's EEG responses (**Fig. 3a**).

Second, we assessed whether these changes in EEG voltage reflect stimulus-related signal. We computed the noise ceiling signal-to-noise ratio (NCSNR) scores[7] for each EEG channel and time point, using the EEG responses for the 102 testing videos. To estimate the noise variance we calculated the variance across the repeated trials for each testing video, and averaged these variances across videos. Next, we estimated the signal variance as the positive half-way rectified difference between the total variance and the noise variance. We then derived the NCSNR scores as the ratio between signal and noise standard deviation. Finally, we plotted the time courses of NCSNR scores averaged across occipital and parietal EEG channels overlaying visual cortex, which showed that EMD's EEG responses contain stimulus-related signal (**Fig. 3b**).

Third, we ran a pairwise decoding analysis on the EEG responses for the 102 testing videos to assess whether a linear classifier can leverage this stimulus-related signal to correctly classify the stimulus conditions of the neural data[8,43,44]. We started by averaging the (up to) 24 repetitions of each video condition into 4 pseudo-trials. Next, we used the pseudo-trial EEG responses of all channels for training and testing linear support vector machines (SVMs) to perform binary classification between each pair of the 102 testing video conditions, following a leave-one-trial-out cross-validation scheme, and independently for each EEG time point. Finally, we averaged the decoding accuracies across all pairwise comparisons, resulting in one time course of decoding accuracy, which showed that stimulus identity is decodable from EMD's EEG responses (**Fig. 3c**).

The results of all three analyses showed common patterns across participants: a first peak at ~0.1 s, a second and higher peak between ~0.2 s and ~0.3 s, followed by a gradual decrease thereafter (despite the stimulus presentation lasting for 3 s), and ending with additional peaks after stimulus offset (i.e., the offset response[12,45–48]). These results show that EMD's EEG responses to videos well encode stimulus-related information, that this stimulus information undergoes several distinct processing stages, and further suggests a difference compared to image-presentation paradigms where the first (ERP, NCSNR, or exemplar decoding) peak at ~0.1 s is typically higher than the following peaks[12,4,13,8].

## EMD's EEG responses exhibit a temporal correspondence with the video stimuli

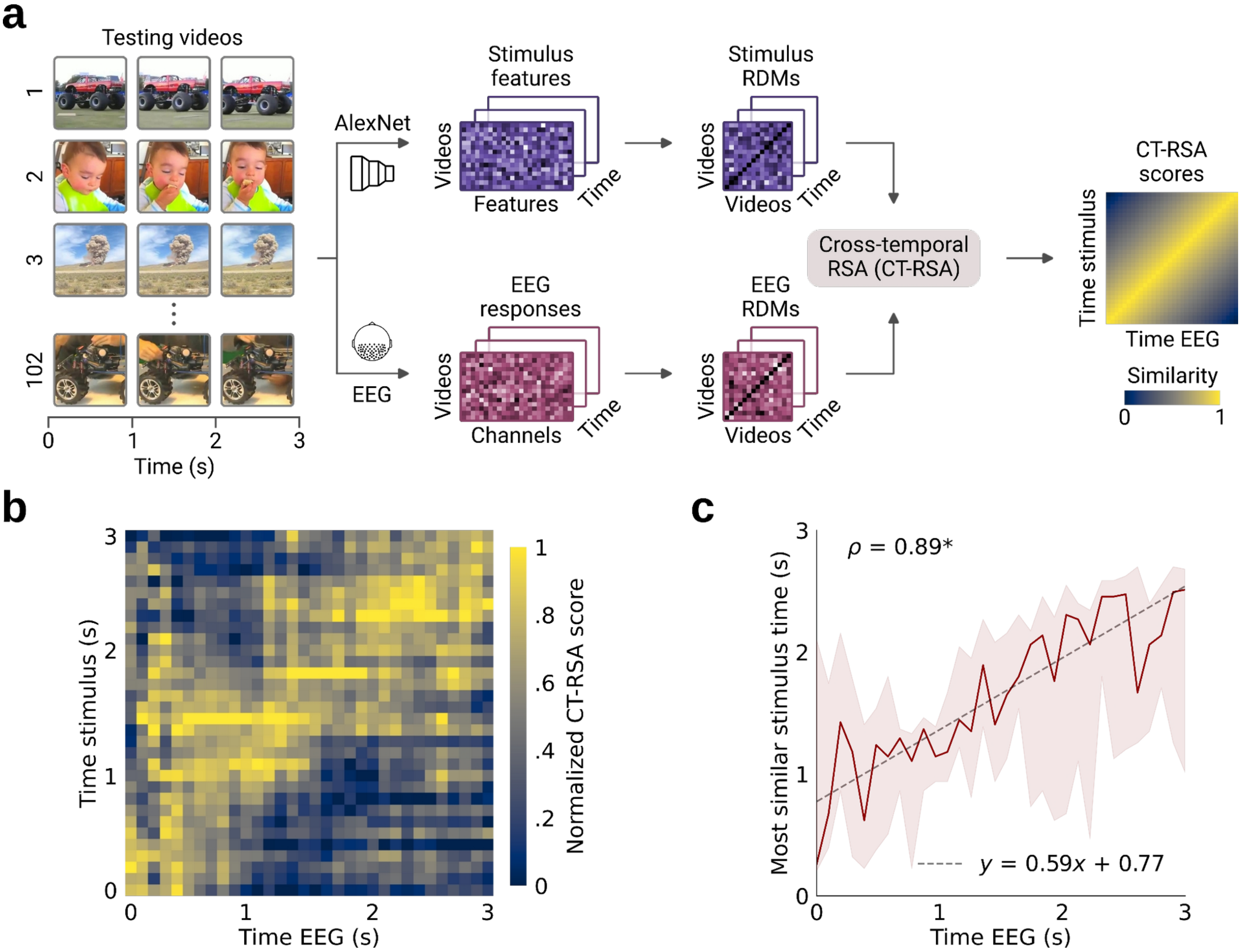

**Fig. 4** EMD's EEG responses exhibit a temporal correspondence with the video stimuli. (**a**) We used cross-temporal representational similarity analysis (CT-RSA) to relate the time courses of the video stimuli with the time courses of EEG occipital and parietal channel responses, using the 102 testing videos. We fed video frames uniformly sampled across the time course of the video stimuli to a pretrained AlexNet, extracted the corresponding stimulus features from the first convolutional layer, and used these features to build representational dissimilarity matrices (RDMs) for each stimulus time point. We adopted an analogous procedure to compute EEG RDMs for each EEG time point. We then applied CT-RSA by correlating the flattened lower triangle of each stimulus feature RDM, with the flattened lower triangle of each EEG RDM. The resulting CT-RSA scores indicate the similarity between each stimulus time point and each EEG time point in terms of representational content. (**b**) 2-dimensional heatmaps of CT-RSA scores between the dynamics of EEG responses (*x* axis) and the dynamics of stimulus features (*y* axis). We normalized the CT-RSA scores in the range [0, 1], independently for each EEG time point, thereby emphasizing the stimulus time point with the highest similarity to each EEG time point. (**c**) For each EEG time point (*x* axis), we computed the most similar stimulus time point (*y* axis) as the average of the 5 stimulus time points with highest CT-RSA scores (red curve). We used a Spearman rank correlation to probe for a monotonically increasing trend between this time course of most similar stimulus time points, and the time point index across the EEG epoch, obtaining a positive significant

relationship of $\rho$ = 0.89 (permutation test, $P$ = 0.0002). To determine a linear relationship, we additionally fit a linear regression between the time point index across the EEG epoch, and the corresponding time course of most similar stimulus time points (dashed gray line). Error margins indicate 95% confidence intervals obtained by bootstrap resampling across participants. (**b-c**) Results reflect participant averages.

Compared to visual neuroscience paradigms based on static images, video stimuli enable the investigation of how dynamic visual stimuli unfold across the EEG time course. Since researchers might want to isolate the neural correlates relative to specific temporally-bounded segments within the full visual events, we asked whether there is a temporal correspondence between the EMD's video stimuli, and the stimulus information encoded in the corresponding EEG responses.

To relate the time courses of the video stimuli with the time courses of EEG responses, we applied cross-temporal representational similarity analysis[15] (CT-RSA; conceptually related to other cross-temporal matching analyses[49,50]) between the stimulus features of 32 frames uniformly sampled across the time course of each of the 102 testing videos, and the corresponding EEG responses for occipital and parietal EEG channels overlaying visual cortex (**Fig. 4a**). To match the EEG time points with the stimulus frames, we averaged the EEG responses into 32 non-overlapping time windows spanning 0-3 seconds with respect to stimulus onset. We used the Net2Brain toolbox[51] to create one Pearson-correlation-based representational dissimilarity matrix (RDM) for each video frame, based on the activations of the first convolutional layer of an AlexNet architecture[52] pretrained on object classification on ImageNet[53]. This yielded 32 RDMs, representing the temporal evolution of stimulus features. Next, we created one Pearson-correlation-based RDM using the repetition-averaged EEG responses of each time point and participant, representing the temporal evolution of information in the EEG. To relate the time course of stimulus features with the time course of EEG responses, we correlated (Spearman's $\rho$) each stimulus feature RDM with each EEG RDM. Finally, we averaged these CT-RSA scores across participants, obtaining a 2-dimensional matrix indicating the representational similarity of each of the 32 stimulus time points with each of the 32 EEG time points.

The resulting CT-RSA matrix had higher scores along the main diagonal (**Fig. 4b**), hinting at a temporal relationship between stimuli and EEG responses. To quantify this relationship, we first assigned each EEG time point to the stimulus time point leading to the highest CT-RSA score, and then used a Spearman rank correlation to probe for a monotonically increasing trend between this time course of most similar stimulus time points and the time point index across the EEG epoch. This resulted in a Spearman correlation score of $\rho$ = 0.89 (**Fig. 4c**), indicating a strong temporal correspondence between the representational dynamics of EMD's video stimuli and EEG responses (permutation test, $P$ = 0.0002). We additionally fit a linear regression between the time point index across the EEG epoch, and the corresponding time course of most similar stimulus time points, which further confirmed the temporal correspondence between video stimuli and EEG responses. Thus, EMD enables fine-grained analyses of how dynamic visual events unfold across the EEG time course.

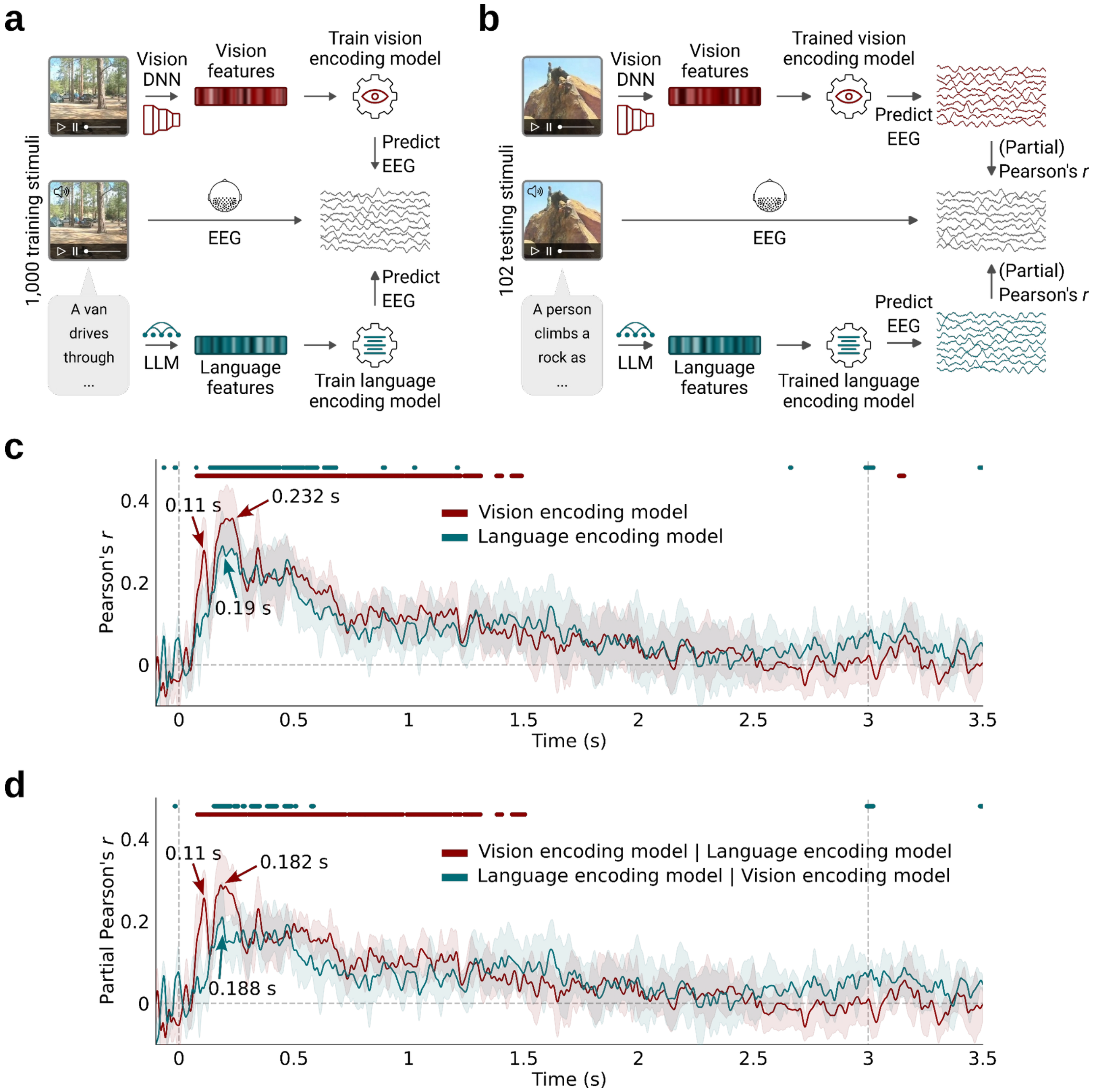

**Fig. 5** EMD's EEG responses are well suited for computational modeling. (**a**) We trained two types of encoding models to predict EEG occipital and parietal channel responses for video stimuli. The first encoding model type predicted EEG responses using vision features obtained by feeding the video stimuli to a vision deep neural network (DNN). The second encoding model type predicted EEG responses using language features obtained by feeding the stimulus text descriptions and spoken transcriptions to a large language model (LLM). (**b**) We tested the prediction accuracy of both types of encoding models using Pearson correlation and partial Pearson correlation analyses. (**c**) Prediction accuracy of vision- and language-feature-based encoding models of EEG responses to video stimuli, measured through a Pearson's correlation between the models' predictions and EMD's EEG responses for the testing video stimuli. (**d**) Partial correlation results for vision- and language-based encoding model predictions, after controlling for the predictions of the other model. (**c-d**) Results reflect participant averages. Arrows denote peak time points. Error margins indicate

95% confidence intervals obtained by bootstrap resampling across participants. Rows of asterisks indicate time points where the correlation or partial correlation scores are significantly greater than 0 (one-sample one-sided $t$-tests, $N = 6$ participants, $P < 0.05$, Benjamini–Hochberg corrected over all time points).

To enable computational brain modeling approaches that rely on separate training and testing data splits for cross-validation, EMD's data is partitioned into a training and a testing split, each with a different number of EEG stimulus repetitions. Thus, in this analysis we showcase EMD's suitability for such modeling through encoding models[14,54,55] trained to map stimulus features onto EEG responses of occipital and parietal EEG channels overlaying visual cortex, and then tested on a left-out portion of data to evaluate their generalization. We assessed two types of commonly used encoding models: one that mapped stimulus features from a vision deep neural network (DNN) onto EEG responses[8], and the other that mapped stimulus features from a large language model (LLM) onto EEG responses[56].

As vision DNN we used S3D[57], a 3D convolutional neural network pretrained on human action classification using videos from Kinetics-400[58]. To obtain the stimulus visual features, we first fed S3D the training and testing video stimuli, and extracted the corresponding layer-wise activations. Next, we used the mean and standard deviation of the training video activations to $z$-score both training and testing video activations. Finally, we reduced the dimensionality of the $z$-scored activations to the number of principal components (PCs) that explained 95% of variance, based on a principal component analysis (PCA) fit on the $z$-scored training video activations. This resulted in training and testing stimulus visual features with 900 PCs. As LLM we used MPNet[59], a sentence-transformer model that maps sentences and paragraphs to a 768 dimensional embedding. To obtain the stimulus language features we fed MPNet the stimulus text descriptions (from EMD's stimulus metadata), and followed the same $z$-scoring and PCA dimensionality reduction procedure described above. This resulted in training and testing stimulus language features with 277 PCs. (Since each video had 5 text descriptions and 1 spoken transcription associated with it, we obtained 6 language feature instances per video. Because averaging these language features across instances would destroy meaningful information idiosyncratic to each instance, we instead retained all 6 instances for the encoding model training and testing.) Finally, we trained ordinary least squares linear regressions that mapped the PCA-reduced vision DNN or LLM features for the training videos onto the repetition-averaged training EEG responses, independently for each channel and time point. (**Fig. 5a;** for the LLM-based models we created 6 copies of EEG responses for each video, so as to match the amount of language feature instances per video, and trained the models with these augmented data).

To assess the prediction accuracy of the two encoding model types, we correlated (Pearson's $r$) their EEG predictions for the vision or language testing stimulus features (we averaged the EEG predictions for the 6 language feature instances per video), with EMD's repetition-averaged testing EEG responses. We computed a correlation score independently for each occipital and parietal EEG channel, and then averaged the correlation scores across channels (**Fig. 5b**). Similarly to the results of the ERP, NCSNR, and pairwise decoding analyses, the prediction accuracy scores of the vision DNN models had a first peak at 0.11 s and a second and higher peak at 0.232 s; instead, the prediction accuracy scores for the LLM models had a single peak at 0.19 s (**Fig. 5c**). For both vision and DNN and language models, this was followed by a gradual decrease towards baseline, with significant prediction accuracies until ~1.5 s and ~0.75 s, respectively (one-sample one-sided $t$-tests, $N = 6$ participants, $P < 0.05$, Benjamini–Hochberg corrected over all time points). Similar patterns persisted in a partial correlation analysis, in which we regressed out the variance associated with one encoding model type from both the measured testing EEG responses and the testing EEG responses predicted by the other encoding model type, and then correlated the resulting residuals so as to isolate the EEG variance uniquely explained by either the vision DNN or LLM model predictions (**Fig. 5d**). Together, these results show that EMD's EEG responses have a rich representational content that is well suited for computational modeling through a diverse set of feature spaces.

EMD’s EEG together with BMD’s fMRI responses enable spatio-temporally resolved analyses of brain responses to videos

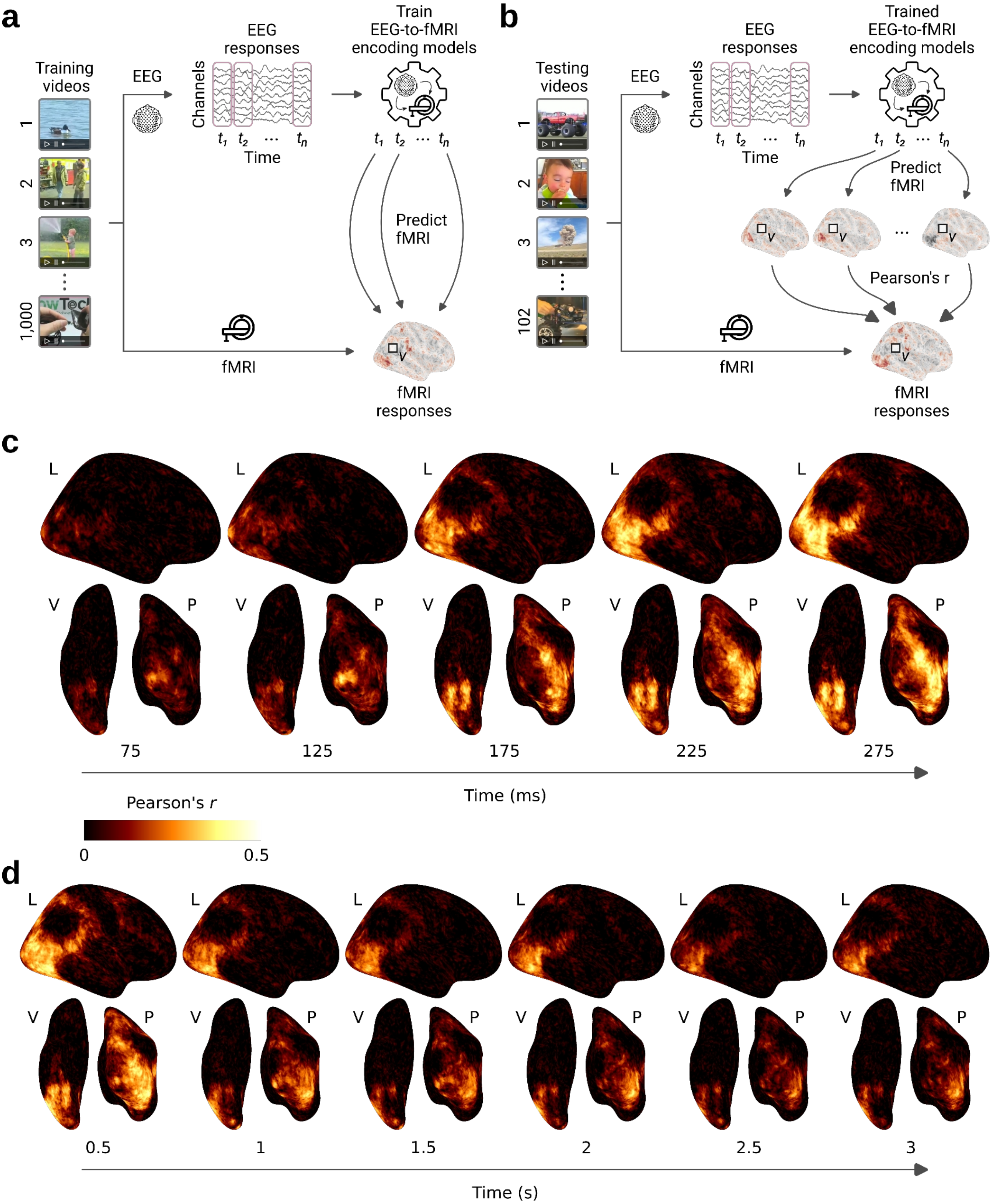

**Fig. 6** EMD’s EEG together with BMD’s fMRI responses enable spatio-temporally resolved analyses of brain responses to videos. (**a**) We trained EEG-to-fMRI encoding models to predict the BOLD Moments Dataset (BMD) fMRI vertex-wise ($v$) responses for the 1,000 testing video stimuli, based on EMD’s EEG channel responses for the same 1,000 videos,

independently for each EEG time point ($t_n$). Each EEG-to-fMRI encoding model is trained using the EEG channel responses aggregated across all 6 EMD participants. (**b**) We used the trained EEG-to-fMRI encoding models to predict time courses of vertex-wise BMD fMRI responses, based on the time course of EMD's EEG channel responses for the 102 testing videos. We then correlated these EEG-predicted fMRI time courses with BMD's measured fMRI responses for the same 102 videos, obtaining one correlation score for each fMRI vertex and EEG time point. (**c**) Time course of correlation scores between EEG-predicted fMRI responses for the 102 testing videos, and BMD's measured fMRI responses for the same videos. Results are shown for the first 275 ms of video presentation, with intervals of 50 ms. (**d**) Time course of correlation scores for the full 3 s of video presentation, with intervals of 0.5 s. (**c-d**) Results reflect averages across all BMD fMRI participants plotted on lateral (L), ventral (V), and posterior (P) views of an inflated right hemisphere in fsaverage space.

EMD complements the BOLD Moments Dataset (BMD)[3] – a large-scale dataset of fMRI responses for the same videos. This enables analyses of human video processing that are resolved in both neural space and time by combining EEG with fMRI data. Thus, in this final analysis we leveraged the high spatial resolution of the BMD fMRI data and the high temporal resolution of the EMD EEG data to achieve a combined spatio-temporally resolved account of human neural responses to videos.

We used an encoding-based framework that leverages the dynamics of EEG to generate time courses of fMRI responses to videos[60,6]. We started by concatenating EMD's repetition-averaged EEG responses of all participants across the channel dimension. Next, we trained ridge regression models that mapped the EEG channel-responses for the 1,000 training videos onto BMD's repetition-averaged fMRI responses for the same training videos (we used BMD's "versionB" preprocessed fMRI data in fsaverage space; note that this fMRI data preparation consists of one beta value per vertex and video stimulus and therefore does not have a temporal dimension). We trained independent models for each fMRI participant and vertex, and for each EEG time point (**Fig. 6a**). We then fed EMD's repetition-averaged, participant-concatenated EEG responses for the 102 testing videos to the trained models to predict time courses of fMRI responses, which we correlated with BMD's repetition-averaged fMRI responses for the same testing videos (**Fig. 6b**). Finally, we averaged these results across fMRI participants, resulting in a correlation score for each fMRI vertex and EEG time point, that is, a spatio-temporally resolved account of linear similarity between the neural responses of each fMRI spatial coordinate and of each EEG temporal coordinate.

In line with the known hierarchical processing of visual cortex[16,19], the correlation scores started rising in early visual cortex during the first 100 ms after video onset, and then propagated to higher processing stages along the ventral, lateral, and dorsal visual streams (**Fig. 6c**), where they lingered for prolonged periods of time (**Fig. 6d**). Together, these results demonstrate that combining EMD's EEG data with BMD's fMRI data enables a spatio-temporally resolved investigation of brain responses to dynamic visual stimuli.

# Data Availability

The EEG Moments Dataset is freely available on OpenNeuro: https://openneuro.org/datasets/ds008257.

# Code Availability

The code used to collect and preprocess EMD, as well as to replicate all results from the technical validation, is available at https://github.com/gifale95/EMD. We additionally provide an interactive Google Colab code tutorial in Python to familiarize with EMD's stimuli and stimulus metadata, as well as with the preprocessed EEG and eye-tracking data, which is available at https://github.com/gifale95/EMD/tree/main/tutorial.

# Author Contributions

A.T.G., P.O., and R.M.C. designed the experiment. A.T.G., P.O., and A.W.Z. collected the data. A.T.G. preprocessed the data. A.T.G., A.W.Z., and C.S. analyzed the data. A.T.G. prepared figures. All authors interpreted results. A.T.G. wrote the manuscript. All authors discussed and edited the manuscript.

# Competing Interests

The authors declare no competing interests.

# Acknowledgments

We thank the HPC Service of FUB-IT, Freie Universität Berlin for computing time (DOI: http://dx.doi.org/10.17169/refubium-26754).

# Funding

R.M.C. is supported by German Research Council (DFG) grants (CI 241/1-3, CI 241/1-7, INST 272/297-1), and European Research Council (ERC) consolidator grant (ERC-CoG-2024101123101). C.S. is supported by an ELLIS Amsterdam Unit grant to I.I.A.G.. A.W.Z. is supported by the University of Amsterdam (UvA) Data Science Centre, as part of the Human Aligned Video AI Lab.